\newcommand{\sandwich}[3]{\mbox{$ \langle #1 | #2 | #3 \rangle $}}

\newcommand{\ket}[1]{\mbox{$ | #1 \rangle $}}
\newcommand{\bra}[1]{\mbox{$ \langle #1 | $}}

\newcommand{\Id}{\mathds{1}}
\newcommand{\beq}{\begin{eqnarray}}
\newcommand{\eeq}{\end{eqnarray}}
\newcommand\abs[1]{\left|#1\right|}

\documentclass[12pt]{iopart}
\renewcommand{\Tr}[1]{\mbox{$\mathrm{Tr}[#1]$}}

\usepackage[T1]{fontenc}
\usepackage[utf8]{inputenc}
\expandafter\let\csname equation*\endcsname\relax
\expandafter\let\csname endequation*\endcsname\relax
\usepackage{amsmath}
\usepackage{amsfonts}
\usepackage{amsthm}
\usepackage{amssymb}
\usepackage{subeqnarray}
\usepackage{dsfont} 
\usepackage{setspace}
\usepackage{graphics}
\usepackage{url}
\usepackage{hyperref}
\usepackage{ifthen}
\usepackage{color}
\usepackage{graphicx}
\usepackage{enumitem}
\usepackage{float}
\usepackage[caption=false]{subfig}

\begin{document}

\title{Measurement-device-independent quantification of entanglement for given Hilbert space dimension}

\author{Koon Tong Goh}
\address{Centre for Quantum Technologies, National University of Singapore, 3 Science Drive 2, Singapore 117543}
\author{Jean-Daniel Bancal}
\address{Centre for Quantum Technologies, National University of Singapore, 3 Science Drive 2, Singapore 117543}
\author{Valerio Scarani}
\address{Centre for Quantum Technologies, National University of Singapore, 3 Science Drive 2, Singapore 117543}
\address{Department of Physics, National University of Singapore, 2 Science Drive 3, Singapore 117542}

\begin{abstract}
We address the question of how much entanglement can be certified from the observed correlations and the knowledge of the Hilbert space dimension of the measured systems. We focus on the case in which both systems are known to be qubits. For several correlations (though not for all), one can certify the same amount of entanglement as with state tomography, but with fewer assumptions, since nothing is assumed about the measurements. We also present security proofs of quantum key distribution without any assumption on the measurements. We discuss how both the amount of entanglement and the security of quantum key distribution (QKD) are affected by the inefficiency of detectors in this scenario.
\end{abstract}

\maketitle

\section{Introduction} 
Entanglement is an essential resource in many quantum information processing and quantum communication applications. Therefore, the certification of entanglement plays an important role in the suite of tests that certify the serviceability of quantum devices.

Most entanglement witnesses discussed in the literature rely on the knowledge of both the Hilbert-space dimension of the systems under study and the measurements being performed. For instance, the criterion $\sigma_x\otimes\sigma_x+\sigma_z\otimes\sigma_z>1$ is an entanglement witness provided the systems are indeed qubits and the measurements are exactly complementary; failure to comply with any of these assumptions may lead to false positives \cite{acin2006,bancal2011,rosset2012}. However, there exist ``device-independent'' (DI) entanglement witnesses: any violation of a Bell inequality\cite{bell1964,brunner2014}, read in the context of quantum theory, certifies the presence of some entanglement under the sole assumption of no-signaling.

Thus, the certification of entanglement depends on the \textit{level of characterization} of the devices. At one extreme, the DI level requires almost no characterization, but is restricted to correlations that violate a Bell inequality and is experimentally demanding since the Bell violation must be loophole-free. The other extreme level, which we call ``tomographic" since it shares the same assumptions as state tomography, is much more versatile and has been routinely implemented for years, at the price of requiring more trust.

\begin{figure}[H]
  \centering
    \includegraphics[width=0.75\textwidth]{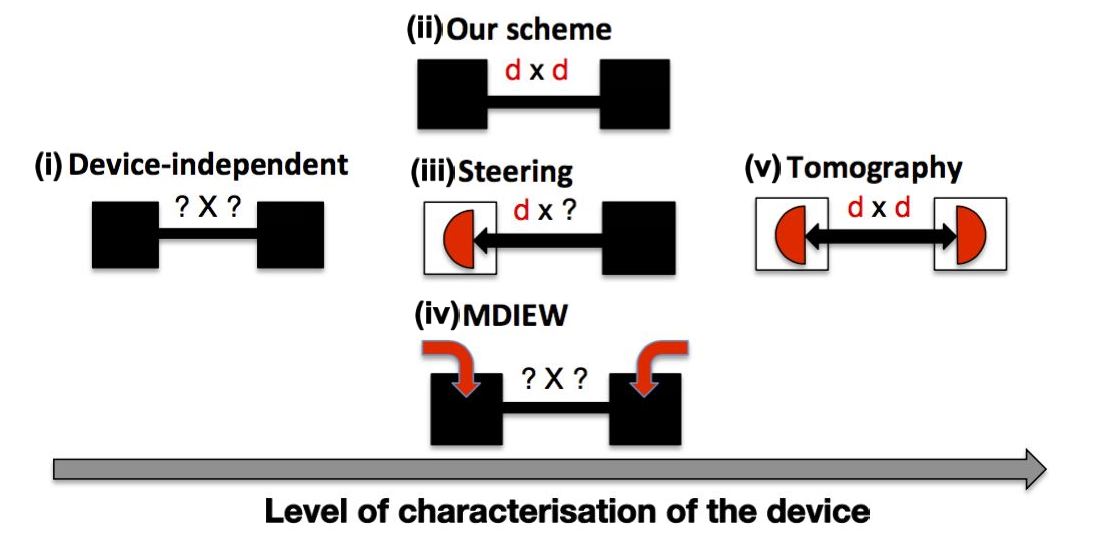}
     \caption{Witnessing entanglement for various levels of characterization of the devices. In the semi-device-independent category one finds: (iv) the MDI Entanglement Witness of \cite{buscemi2012,branciard2013}, that requires the use of known local quantum states instead of classical inputs; (iii) steering \cite{wiseman2007}, in which Alice's box is tomographically known while Bob's box is fully unknown; (ii) the case of known dimensions considered here. Besides (ii), Ref.~\cite{moroder2012} introduced further layers of assumptions on the measurements leading all the way to tomography; ref.~\cite{moroder2014} studied two black boxes with dimensions $d\times ?$.}
     \label{fig1}
\end{figure}

Between these two extremes, various \textit{semi-device-independent} levels can be defined by relaxing some of the assumptions (Figure \ref{fig1}). This paper deals with one such relaxation: the dimensionality of the systems is trusted but the measurements are not. Entanglement certification with known dimension was first introduced by Moroder and Gittsovich \cite{moroder2012}, who discussed how to certify the presence of entanglement and derived analytical conditions for some cases. We describe certifiable lower bounds on the \textit{amount} of entanglement. For the case of bipartite qubits, we show that in several relevant cases one certifies as much entanglement (in terms of concurrence) as with tomography, but with fewer assumptions. 

In terms of entanglement being a resource for quantum key distribution (QKD), we derive security bounds for implementations of the BB84 and the six-state protocols that use entangled qubits. Similar work was done for the BB84 protocol in~\cite{gittsovich2013}, as well as in the recent, independent work~\cite{woodhead2015}.
This work fits in the broad class of ``measurement-device-independent'' (MDI) approaches, in which all the imperfections of the measurement devices and detectors don't need to be modelled. The most famous MDI scheme is one for QKD, that does not require entanglement but rather post-selects it \cite{lo2012,braunstein2012}. A different MDI scheme for entanglement certification \cite{buscemi2012,branciard2013} requires a modification of the usual setups \cite{verbanis2015}: measurement settings are not chosen with classical inputs but with local quantum states; the latter must then be well characterized.

We finally stress that our work is different from the series of works devoted to ``dimension witnesses'', where the main goal is to establish that some observed statistics cannot be obtained by measuring low-dimensional systems (whether entangled or not). The dimension witnesses that are also Bell inequalities definitely certify entanglement, but as should be clear from the discussion above, this is not what we are aiming at. Rather, one of our goals is to certify entanglement with statistics that don't violate any Bell inequality, thanks to the additional dimensional constraints \footnote{In particular, the recent work of Navascu\'es and coworkers \cite{Navascues2014} fits in this category: in the examples where they claim to certify entanglement, it's because they are using a Bell inequality. That being said, it can't be excluded that some of the tools developed there could find a broader application.}.

\section{General framework}

 We want to find how much entanglement (if any) is needed to reproduce a set of measurement statistics, $p := p(ab|xy)$, under the constraint that $\rho \in{\cal L}\big(\mathcal{H}^d\otimes \mathcal{H}^{d'}\big)$ for fixed $d$ and $d'$. Local uncorrelated ancillas are free resources, i.e. the local measurements can be POVMs. However, classical shared randomness is not a free resource: all correlations must be accounted for by the allowed dimensions \footnote{If we were to fix the dimensions but leave classical shared randomness free, only statistics that violate Bell inequalities would certify entanglement, thus this level of characterisation would not detect entanglement where the fully DI level doesn't. Previous works have considered the question of quantifying entanglement in this context~\cite{Verstraete2002,Liang2011}.}.

In general, the amount of entanglement certifiable by observing measurement statistics $p$ for a fixed dimension of the Hilbert space, denoted by $e(p)$, is given by:

\begin{align}\label{optgen}
e(p) := \min_{\rho, \Pi^a_x,\Pi^b_y} & E(\rho) \\
s.t. \;\; & p(a,b|x,y) = \Tr{\rho\, \Pi^a_x \otimes \Pi^b_y} \;\; \forall x,y,a\; \& \;b, \nonumber \\
\;\; & \sum_a \Pi^a_x =  \sum_b \Pi^b_y = \Id \;\; \forall x\;\&\;y, \nonumber \\
\;\; & \Pi^a_x , \Pi^b_y \geq 0 \;\; \forall x,y,a\; \& \;b , \nonumber \\
\;\; & \rho \in  {\cal L}\big(\mathcal{H}^d\otimes \mathcal{H}^{d'}\big)\, \nonumber
\end{align}
where $E(\rho)$ is an entanglement monotone of the bipartite state $\rho$, such as the negativity \cite{vidal2002} for any arbitrary $d$ and $d'$ or the concurrence \cite{wootters1998} for the case of $d=d'=2$. Here, the optimization runs over all states $\rho$ and measurements $\Pi^a_x$ and $\Pi^b_y$ which are compatible with the statistics $p(a,b|x,y)$.

The structure of the problem resembles many optimisation problems encountered in this field. As it often happens, the number of parameters is large enough to make analytical solutions cumbersome: even for the simpler problem of certifying the existence of entanglement, without quantification, Moroder and Gittsovich could solve explicitly only few-parameter classes \cite{moroder2012}. Moreover, since the set of quantum statistics for fixed dimension is not convex \cite{pal2009,donohue2015}, the optimisation is not a semi-definite program. Fortunately, heuristic numerical optimisations are reliable if the number of parameters is not too large. In the following section, we will show explicit solutions of this problem for $d=d'=2$. Higher dimensional cases can be addressed similarly by solving (1) for the negativity.

\section{Explicit examples on bipartite qubits states}

For the rest of the paper, we focus on two-qubit states. We base our figure of merit on concurrence $C(\rho)$ \cite{wootters1998} which is defined by:
\begin{equation}
C(\rho)=\max(0,e_1-e_2-e_3-e_4)
\end{equation}
where $e_1$,$e_2$,$e_3$ and $e_4$ are the eigenvalues of $\sqrt{\sqrt{\rho}\tilde{\rho}\sqrt{\rho}}$ such that $e_1\geq e_2 \geq e_3 \geq e_4$ and $\tilde{\rho}=(\sigma_y \otimes \sigma_y)\rho(\sigma_y \otimes \sigma_y)$. It is well known that the concurrence is an entanglement monotone that satisfies $0\leq C(\rho) \leq 1$, $C(\rho)=0$ iff $\rho$ is separable, and $C(\rho)=1$ iff $\rho$ is maximally entangled. It is also related to the entanglement of formation by $E_F(\rho)=h\big(\frac{1}{2}(1-\sqrt{1-C^2(\rho)})\big)$ where $h$ is binary entropy \cite{wootters1998}.

The task is now to solve the optimisation \eqref{optgen} for $d=d'=2$ and $E(\rho)=C(\rho)$. In the following sections, we describe the result of this optimization for several choices of statistics $p$.

\subsection{Well-known correlations from dichotomic measurements}

We consider first some statistics with binary outcomes $a,b \in \{-1,+1\}$. Specifically, we consider three families that can be obtained by suitable projective measurements on the two-qubit state unitarily equivalent to the Werner state $\rho_W = W \ket{\Phi^+}\bra{\Phi^+} + \frac{1-W}{4}\Id$, where $\ket{\Phi^+} = \frac{1}{\sqrt{2}}(\ket{00} + \ket{11})$:
\begin{itemize}
\item The \textit{CHSH family} is defined as
\begin{equation}
p_{\text{CHSH}}(a,b|x,y)=\frac{2+ab(-1)^{xy}\sqrt{2}W}{8},x,y\in\{0,1\}\,.\label{pchsh}
\end{equation}
These statistics are achievable from $\rho_W$ with $\Pi_{x=0}^a=\frac{1}{2}(1+a\sigma_z)$, $\Pi_{x=1}^a=\frac{1}{2}(1+a\sigma_x)$, $\Pi_{y=0}^b=\frac{1}{2}(1+b\frac{\sigma_z+\sigma_x}{\sqrt{2}})$ and $\Pi_{y=1}^b=\frac{1}{2}(1+b\frac{\sigma_z-\sigma_x}{\sqrt{2}})$. These statistics violate the CHSH Bell-type inequality for $W>\frac{1}{\sqrt{2}}$: in that range, entanglement can be certified in a DI way.
\item The \textit{BB84 family} is defined as
\begin{equation}
p_{\text{BB84}}(a,b|x,y)=\frac{1+ab\delta_{x,y}W}{4}\,,x,y\in\{0,1\}\,,\label{pbb84}
\end{equation}
achievable from $\rho_W$ with $\Pi_{x=0}^a=\frac{1}{2}(1+a\sigma_z)$, $\Pi_{x=1}^a=\frac{1}{2}(1+a\sigma_x)$, $\Pi_{y=0}^b=\frac{1}{2}(1+b\sigma_z)$ and $\Pi_{y=1}^b=\frac{1}{2}(1+b\sigma_x)$. These statistics can be obtained with shared randomness of dimension 4, therefore no DI certification of entanglement is possible.
\item The \textit{six-state family}
\begin{equation}
p_{\text{six-states}}(a,b|x,y)=\frac{1+ab(-1)^{\delta_{x,2}}\delta_{x,y}W}{4},x,y\in\{0,1,2\}\,
\end{equation}
achievable from $\rho_W$ with the same measurements as the previous ones plus $\Pi_{x=2}^a=\frac{1}{2}(1+a\sigma_y)$ and  $\Pi_{y=2}^b=\frac{1}{2}(1+b\sigma_y)$. These statistics can be obtained with shared randomness of dimension 8, therefore again no DI certification of entanglement is possible.
\end{itemize}

Let us now see what can be said when these statistics are supplemented with the assumption that the measured state is a two-qubit state. Before turning to our approach, we check sufficient criteria for the \textit{existence} of entanglement from Ref.~\cite{moroder2012}. Concretely, when there are two measurements per party and unbiased marginals, as in $p_{\text{CHSH}}$ and $p_{\text{BB84}}$, one can use item (4) in Proposition 2 of Ref.~\cite{moroder2012}: entanglement is certified if $\sqrt{\lambda_1}+\sqrt{\lambda_2}>\sqrt{2}$, where the $\lambda$'s are the singular values of the $2\times 2$ correlation matrix $D_2$  whose entries are $[D_2]_{xy}=\langle A_xB_y\rangle$ with $A_x=\sum_{a=\pm 1} a\Pi_{a}^x$, $B_y=\sum_{b=\pm 1} b\Pi_{b}^y$. For both the CHSH and the BB84 family, this criterion certifies entanglement when $W>\frac{1}{2}$. For $p_{\text{six-states}}$, that has three measurements per party, one can use Eq.~(46) of Ref.~\cite{moroder2012} for $n=3$ and $d=2$: for unbiased marginals, this criterion reads $|det(D_3)|>\frac{1}{27}$ with $D_3$ the $3\times 3$ correlations matrix again defined by $[D_3]_{xy}=\langle A_xB_y\rangle$. The result is that $p_{\text{six-states}}$ certifies qubit entanglement for $W>\frac{1}{3}$. Given that $\rho_W$ is separable for $W\leq\frac{1}{3}$, this is also a necessary condition: in other words, $p_{\text{six-states}}$ detects the whole range of Werner state entanglement without knowing the measurements.

\begin{figure}[H]
  \centering
    \includegraphics[width=0.6\textwidth]{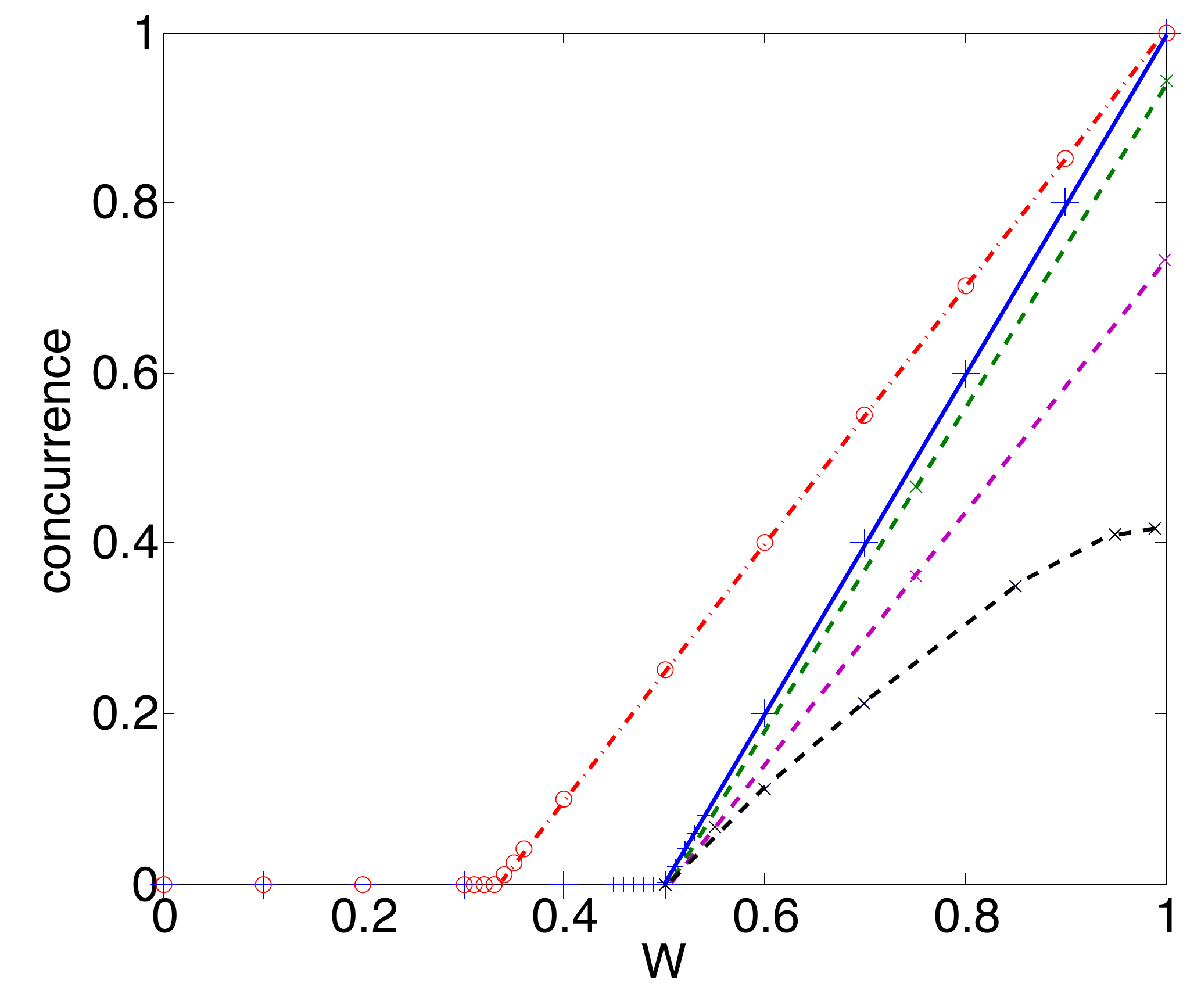}
     \caption{(Color online) Plot of $c(p)$ against $W$. We find $c(p_{\text{CHSH}})=c(p_{\text{BB84}})$ (blue ``$+$'' data points) and $c(p_{\text{six-states}}) =C(\rho_W)$ (red ``$O$'' data points); the blue solid and red dot-dashed lines are the best-fit lines. The green, magenta and black dotted lines on ``$x$'' data points are guide for the eyes of $c(p_{\text{BB84}}^{\text{noisy}})$ for $\epsilon_A=\epsilon_B=\frac{2}{3}$, $\frac{1}{3}$ and $\frac{1}{10}$ respectively.}
     \label{plot1}
\end{figure}

We turn now to our approach, which not only certifies the existence of entanglement, but puts a lower bound on its amount. In~\ref{sec:StateConstraints} we describe the constraint that the relation $p(a,b|x,y)=\Tr{\rho \cdot \Pi^a_x \otimes \Pi^b_y}$ imposes on the two-qubit state $\rho$ for $p_{\text{BB84}}$ and $p_{\text{six-states}}$. The optimization~\eqref{optgen} is the performed by minimizing
$C(\rho)$ numerically over the remaining free parameters. The result is plotted in Figure \ref{plot1}. We observe that $c(p_{\text{CHSH}})=c(p_{\text{BB84}})$ over all the range of $W$: the fact of using a Bell inequality does not provide any advantage in this example. Also, we find that $c(p_{\text{six-states}})=C(\rho_W)$: moving to the tomographic level would not improve the certification if the state was actually $\rho_W$.

A last remark on these three families: for $W=1$, it is known that $p_{CHSH}$ provides a DI \textit{self-testing} of $\ket{\Phi^+}$ and the corresponding measurements \cite{popescu1992}. The constraints we found readily imply that, if one adds the qubit assumption, also $p_{\text{BB84}}$ and $p_{\text{six-states}}$ self-test $\ket{\Phi^+}$ and the corresponding measurements (see~\ref{sec:StateConstraints} for an explicit proof).

Having warmed up with what are arguably the most studied families of correlations, we turn to use our method on three other examples.

\subsection{A slice in the CHSH polytope}

\begin{figure}[H]
  \centering
    \includegraphics[width=0.55\textwidth]{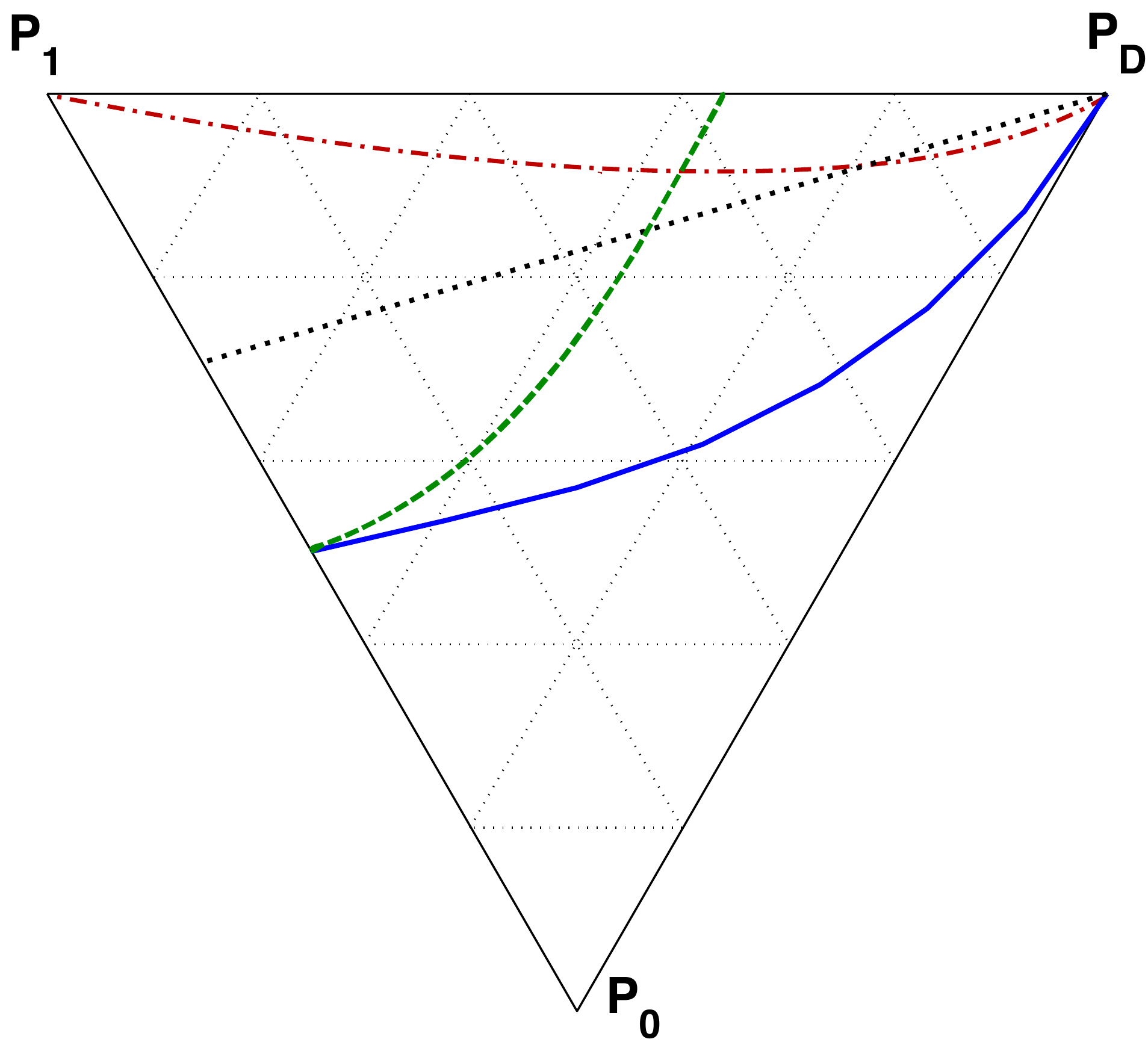}
     \caption{(Color online) A two-dimensional slice in the 8-dimensional CHSH probability space. $P_0$ and $P_{1}$ represent the statistics $p_{\text{CHSH}}$ for $W=0$ and $W=1$ respectively, while $P_D$ is the deterministic point defined by $p(a=+1|x)=p(b=+1|y)=1$ $\forall x,y$. The black dotted line is the CHSH facet: points above that line certify entanglement in a DI way. The red dash-dotted curve, drawn with the equation
provided in the caption of figure 3 of \cite{donohue2015}, bounds the set of quantum correlations that can be achieved with two qubits: this set is clearly not convex and even some classical correlations are excluded. Our calculation provides the blue solid line: points above that line certify entanglement if the system is two qubits. On the line joining $P_0$ and $P_1$, representing the family $p_{\text{CHSH}}$, entanglement is certified for $W>\frac{1}{2}$, as proved already in Fig.~\ref{plot1}. Eq.~(46) of \cite{moroder2012} certifies the existence of entanglement only for points above the green dashed curve if the system is two qubits.}
     \label{ternplot}
\end{figure}

The first example remains with two measurements and binary outcomes: in the 8-dimensional CHSH polytope, we consider the two-dimensional triangle bounded by the family $p_{\text{CHSH}}$ and by a deterministic point (in particular then, all the probability points but the $p_{\text{CHSH}}$ line have biased marginals). Other properties in this triangle were studied in Ref.~\cite{donohue2015}. Our result for the existence of entanglement (Figure \ref{ternplot}) shows more clearly the difference with DI certification and the improvement over Eq.~(46) of \cite{moroder2012}.

\subsection{A SIC-POVM}

For the last example, we assume that both Alice and Bob perform a single measurement, that of a four-outcome symmetric, informationally complete, positive operator valued measure (SIC-POVM) \cite{renes2004}. Concretely, let us use the SIC-POVM defined by
\begin{align}
\Sigma_{0}&=
\begin{pmatrix}
\frac{1}{2} & 0 \\
0 & 0 
\end{pmatrix}
\;,\; \Sigma_{1}=
\begin{pmatrix}
\frac{1}{6} & \frac{1}{3\sqrt{2}}\\
 \frac{1}{3\sqrt{2}} &  \frac{1}{3} 
\end{pmatrix} \label{sic}\\
\Sigma_{2}&=
\begin{pmatrix}
 \frac{1}{6} &  \frac{\chi}{3\sqrt{2}} \\
\frac{\chi^*}{3\sqrt{2}}  & \frac{1}{3} 
\end{pmatrix}
\;,\; \Sigma_{3}=
\begin{pmatrix}
 \frac{1}{6} &  \frac{\chi^*}{3\sqrt{2}} \\
\frac{\chi}{3\sqrt{2}}  & \frac{1}{3} 
\end{pmatrix} \nonumber
\end{align}
where $\chi=e^{\frac{2\pi i}{3}}$.

We consider the statistics $p_S(\theta)=\sandwich{\psi(\theta)}{\Sigma_{a}\otimes\Sigma_{b}}{\psi(\theta)}$ arising when both Alice and Bob measure the SIC-POVM on their half of a pure non-maximally entangled states of two qubits $\ket{\psi(\theta)}= \cos(\theta)\ket{00} + \sin(\theta)\ket{11}$, with $\theta\in [0,\pi/4]$. On the one hand, by definition of a SIC-POVM, at the tomographic level these measurement statistics can be used to reconstruct the state exactly: thus, entanglement is certified for all $\theta>0$. On the other hand, since there is only one measurement per party, there is no hope of any DI certification of entanglement.

\begin{figure}[H]
  \centering
    \includegraphics[width=0.6\textwidth]{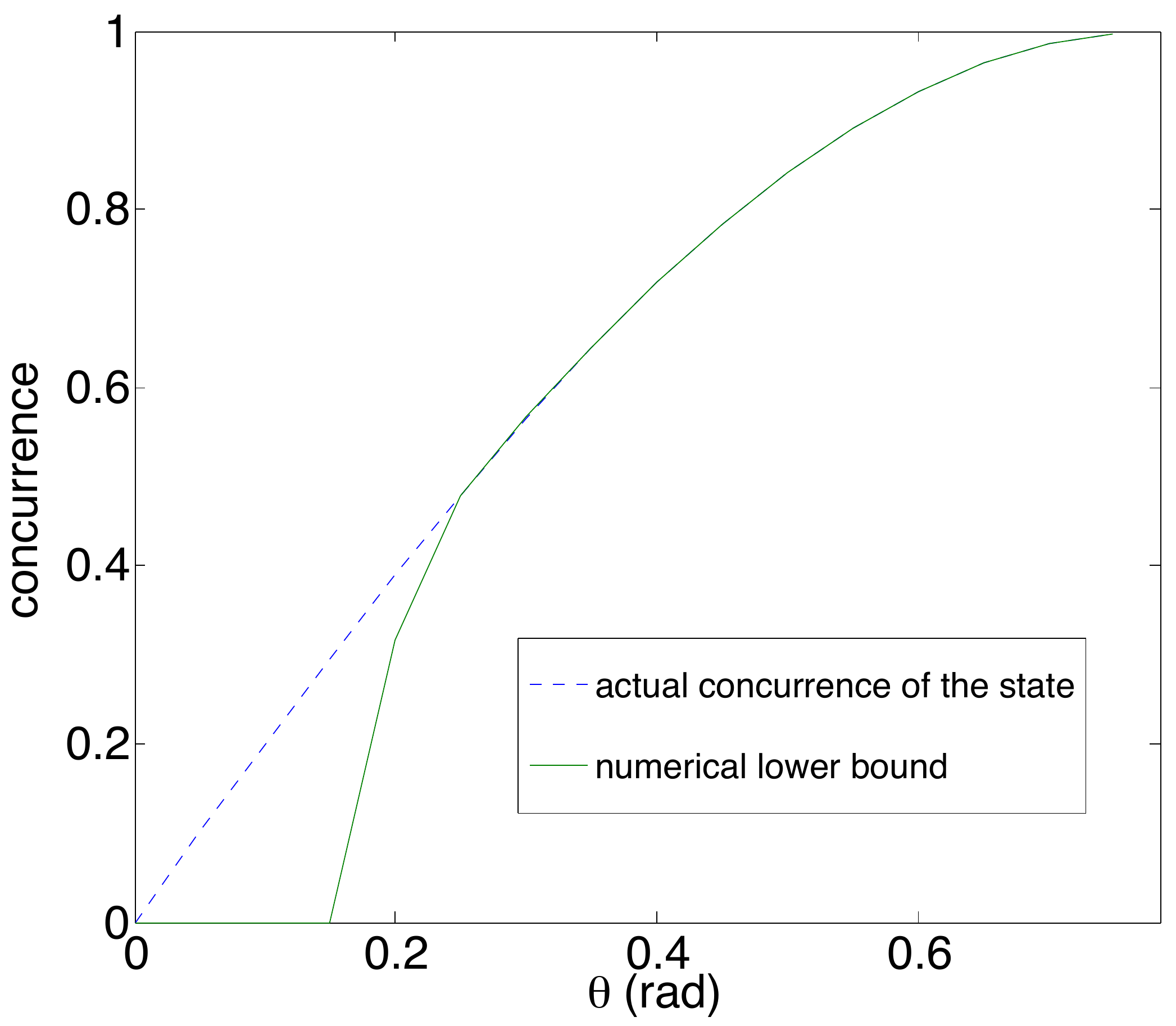}
     \caption{(Color online) Measurement of the four-outcome SIC-POVM by both Alice and Bob: certifiable two-qubit entanglement $c(p_S)$ (solid line) and concurrence of $\ket{\psi(\theta)}$ (dotted line).}
     \label{sicpovmplot}
\end{figure}

If we assume only the dimension, the statistics do certify entanglement for a large range of $\theta$, but become achievable with a POVM on separable states for small values of $\theta$ (see Figure \ref{sicpovmplot}). Thus, the family $p_S(\theta)$ is such that tomography certifies more entanglement than the sole assumption of qubits, thus adding to similar examples in Section III E of Ref.~\cite{moroder2012}.

\subsection{Detection loophole for entanglement quantification}

The last example of statistics that we study deals with the \textit{detection loophole}. Typically mentioned in the context of DI Bell experiments, the loophole can appear at any level of characterization of the devices. Concretely, it was noticed in 2007 by Skwara and co-workers that it may actually apply to entanglement certification at the tomographic level \cite{skwara2007}. In their study, they considered detecting entanglement by the optimal entanglement witness in the absence of losses; then added the conservative assumption that all the non-detection events were hiding actual outcomes that would have contributed negatively to the entanglement witnessing. In this framework, they found sufficient criteria to detect entanglement and conjectured that it would be impossible to certify entanglement if Alice's and Bob's detection efficiencies were $\epsilon_A=\epsilon_B<\frac{2}{3}$.

In our formalism, the study of the detection loophole for entanglement certification is much more direct: we simply treat the no-detection event as another outcome (as we'll highlight later, things are subtler for adversarial scenarios like cryptography).

We base our case study on the BB84 family $p_{\text{BB84}}$, assuming that we would observe those statistics with perfect detectors. The third outcome 0 is assigned to the events in which neither detector clicked: now $a, b\in\{-1,+1,0\}$. If now both Alice's detector have efficiency $\epsilon_A$ and both Bob's detectors efficiency $\epsilon_B$, and the firing of a detector is an uncorrelated process, the observed three-outcome statistics would be
\begin{align}
p^{\text{noisy}}_{\text{BB84}} &= \abs{ab} \epsilon_A\epsilon_B p_{\text{BB84}} + \delta_{a,0}\delta_{b,0}(1-\epsilon_A)(1-\epsilon_B) \\
&+\abs{\abs{ab}-1}\abs{\delta_{a,0}\delta_{b,0}-1}\frac{1}{2}[\epsilon_A(1-\epsilon_B)+(1-\epsilon_A)\epsilon_B]\,. \nonumber
\end{align}

As before, we now assume that these statistics have actually been observed, without any characterisation of the detectors, nor any information about the physical process that generated them apart from the fact that it was a three-outcome POVM on qubits. The resulting lower-bound on the concurrence $c(p)$ is plotted in Fig.~\ref{plot1} for some choices of $\epsilon_A=\epsilon_B$. The result shows that the \textit{amount} of entanglement that can be certified decreases with the efficiency, but the \textit{presence} of entanglement is certified irrespective of the efficiency (numerically proved for $\epsilon\geq\frac{1}{10}$) if the dimensionality is known. 



\section{Applications to quantum key distribution}

The first unconditional security proofs of QKD protocols were obtained by modeling the signals as qubits and by assuming the best-suited measurements (for both BB84 and six-states, complementary bases). We work at this elementary level to show that very similar performances can be certified even if the knowledge about the measurements is removed. We are of course aware that practical security proofs have gone a long way towards a more realistic modeling of both the signals and the measurements \cite{scarani2009, coles2015}: as a future project, it will be interesting to see if the approach started here can also be used to remove assumptions from these more sophisticated descriptions.

\subsection{Framework}

In order to make this text readable, we need to explain quickly how the security bound is found (we work with the same tools as Appendix A of \cite{scarani2009}). The goal of a security proof is to estimate the information that the eavesdropper may have. Asymptotically, Eve's information per qubit is given by $\chi(A:E)=S(\rho_E)-\sum_{a}p(a)S(\rho_E^{a})$ where $a$ are the outcomes of Alice's key measurement, $S(\cdot)$ denotes von Neumann entropy. This holds for the most powerful Eve, one that holds a purification of $\rho_{AB}$. Thus, given $\rho_{AB}$, one has to write down a purification $\ket{\psi}_{ABE}$, then trace Bob out and compute $\chi(A:E)$, this latter step requiring the knowledge of what Alice's key measurements are. The parameters of the protocol may not determine $\rho_{AB}$ fully: one may have to maximise Eve's information over all states compatible with the observation. Ultimately, the figure of merit is the secret key fraction $r=1-h(Q)-\chi(A:E)$ where $h(Q)$ is binary entropy and $Q$ is the Quantum Bit Error Rate (QBER) in the key bits.

If one knows only the dimension, the constraints on the state are given by $p$ and we don't know anything about which measurement is actually performed to obtain the key. The certifiable secret key fraction is then be given by the optimisation
\begin{align}
r' = \min_{\rho_{\text{AB}}, \Pi^a_x, \Pi^b_y} & 1-h(Q)-\chi(A:E) \label{crypto} \\
s.t. \;\; & p = \Tr{\rho_{\text{AB}} \cdot \Pi^a_x \otimes \Pi^b_y} \;\; \forall x,y,a\; \& \;b, \nonumber \\
\;\; & \sum_a \Pi^a_x =  \sum_b \Pi^b_y = \Id \;\; \forall x\;\&\;y, \nonumber \\
\;\; & \Pi^a_x , \Pi^b_y \geq 0 \;\; \forall x,y,a\; \& \;b , \nonumber \\
\;\; & \left|\psi\right>_{\text{ABE}} = \sum_{j}\sqrt{\lambda_j} \left|\phi_j\right>_{\text{AB}}\left|j\right>_{\text{E}} \nonumber \\
\;\; & \rho_E = \text{Tr}_{\text{AB}}\left[\left|\psi\right>\left<\psi\right|_{\text{ABE}} \right]\nonumber \\
\;\; & \rho^{a=i}_E = \frac{\text{Tr}_{\text{AB}}\left[\sqrt{\Pi^{a=i}_x\otimes\Id}\left|\psi\right>\left<\psi\right|_{\text{ABE}} (\sqrt{\Pi^{a=i}_x\otimes\Id})^\dagger\right]}{\text{Tr}\left[\sqrt{\Pi^{a=i}_x\otimes\Id}\left|\psi\right>\left<\psi\right|_{\text{ABE}} (\sqrt{\Pi^{a=i}_x\otimes\Id})^\dagger\right]}\nonumber \\
\;\; & \rho_{\text{AB}} \in  {\cal L}\big(\mathcal{H}^2\otimes \mathcal{H}^2\big) \nonumber
\end{align} where $\lambda_j$ and $\left|\phi_j\right>_{\text{AB}}$ denotes the eigenvalues and eigenstates of $\rho_{\text{AB}}$ respectively. It is customary to assume that the key is obtained from the measurements $A_0$ and $B_0$, whence the QBER is given by $Q=p(a\neq b|0,0)$.

\subsection{Application to the BB84 and six-states protocols}

At the tomographic level of characterization, the six-state protocol is tomographically complete, so one actually knows $\rho_{AB}$; in BB84, there remains one free parameter over which to optimize. If the error fractions are the same in all bases, for BB84 one obtains the well-known result
\begin{align}
r_{\text{BB84}}&=1-2h(Q)
\end{align} while the the six-state protocol has the slightly higher yield
\begin{align}\label{sixstates}
r_{\text{six-states}}= 1-\frac{3Q}{2}\log_2{\left(\frac{Q}{2}\right)}-\left(1-\frac{3Q}{2}\right)\log_2{\left(1-\frac{3Q}{2}\right)}\,.
\end{align}

\begin{figure}[H]
  \centering
    \includegraphics[width=0.6\textwidth]{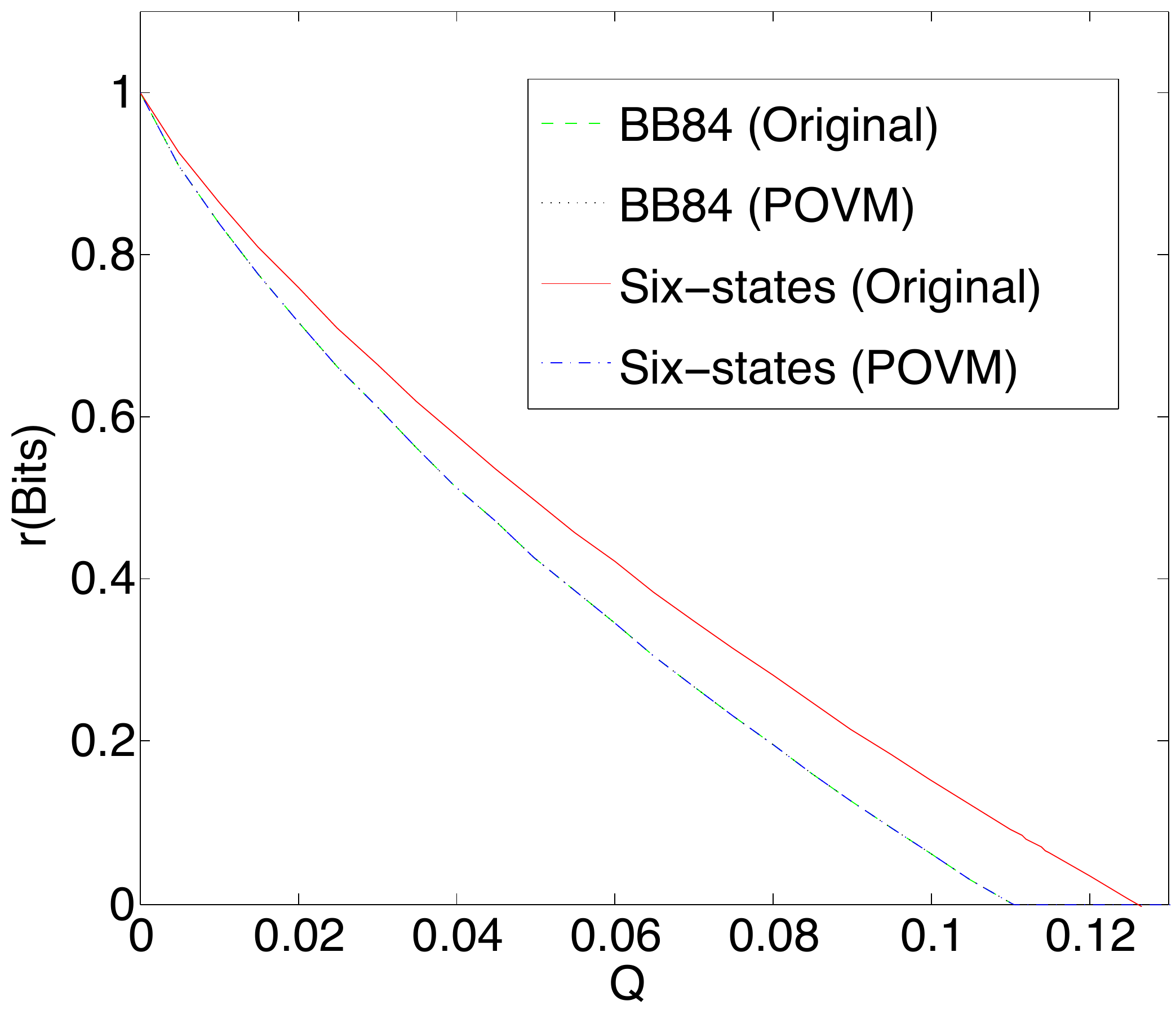}
     \caption{Secret key fraction against $Q$. For both the BB84 and the six-states protocols, the results of optimisation \eqref{crypto} yield $r'_{\text{BB84}}=r'_{\text{six-states}}=r_{\text{BB84}}$; the slightly higher curve is the tomographic yield of the six-state protocol \eqref{sixstates}.}
     \label{plotqkd}
\end{figure}

Figure \ref{plotqkd} shows the results of optimisation \eqref{crypto} for the BB84 and six-states protocols, for the observed statistics $p_{\text{BB84}}$ and $p_{\text{six-states}}$ respectively (see~\ref{sec:cryptoAppendix} for the constraints that they imply; notice that $Q=\frac{1-W}{2}$). We find $r'_{\text{BB84}}=r_{\text{BB84}}$: the relaxation of assumption on measurement does not affect the secret key fraction in the BB84 protocol. This result had been obtained analytically thanks to the special properties of $p_{\text{BB84}}$ \cite{gittsovich2013} and has been recently re-derived in a parallel work \cite{woodhead2015}. These analytical derivations are possible thanks to the high symmetry of the BB84 protocol and can in fact be made with by assuming only that Alice measures a qubit, while Bob's system has arbitrary dimension.

For the six-states protocol, the relaxation of the assumption on the measurements brings the secret key fraction down to match that of BB84: $r'_{\text{six-states}}=r_{\text{BB84}}$. In this case, inspection shows that the optimum is achieved by measuring the state $\rho_{\text{AB}}=\frac{1}{4}\big(\Id\otimes\Id + W(\sigma_x\otimes \sigma_x-\sigma_y\otimes \sigma_y)+ \sigma_z \otimes \sigma_z\big)$ with $\hat{A}_0=\sigma_z$, $\hat{A_1}=\hat{B}_1=\sigma_x$, $\hat{A_2}=\hat{B}_2=\sigma_y$, and $B_0$ given by the two-outcome POVM $\Pi^{b=\pm 1}_{y=0}=\frac{\Id\pm W\sigma_z}{2}$.

A thorough study of the effect of imperfect detection in QKD goes beyond the scope of this paper. Indeed, in the presence of losses in the channel, the signals can't be assumed to be qubits, as we are doing here: there should be at least a third state, representing the leakage. Besides, BB84 is by definition a protocol with binary output, so we can't use the three-outcomes $p^{\text{noisy}}_{\text{BB84}}$ as such. One should either define a modified protocol, or process the data to obtain two outcomes in an optimal way. For a preliminary study, we consider the two-outcome statistics obtained from $p^{\text{noisy}}_{\text{BB84}}$ upon sending all the `$0$' to `$+1$' (see~\ref{sec:cryptoAppendix} for details). The main result is that, for $W=1$, $r=0$ for $\epsilon\lesssim 87\%$. Above this efficiency, for a given $Q$ it is possible to find a key rate $r$ slightly larger than $r_{\text{BB84}}$, an effect also seen in standard security proofs when local noise is added \cite{kraus2005}.

\section{Conclusion}
We study how much entanglement can be certified when knowing the statistics and the dimension of the Hilbert space of the measured systems, but nothing about the measurements. For the case of two qubits, the amount that can be certified is similar, and in several cases identical, to that which can be certified by tomography, but with fewer assumptions. We also showed how to apply the same level of characterisation in the simplest level of security proofs in quantum key distribution. In the case considered here, the detection inefficiency does not affect the certification of the presence of entanglement, but does affect the more quantitative certifications.

\ack We thank Nicolas Brunner, Kai Chen, Patrick Coles, John Donohue, Nicolas Gisin, Marcus Huber, Norbert L\"utkenhaus, Stefano Pironio, Elie Wolfe, Erik Woodhead, Yizheng Zhen and Yulin Zheng for discussions and comments.

This work is funded by the Singapore Ministry of Education (partly through the Academic Research Fund Tier 3 MOE2012-T3-1-009) and by the National Research Foundation of Singapore.

\section*{References}
\bibliographystyle{iopart-num}
\bibliography{bib}

\newpage
\appendix
\section{Supplementary Information: Constraints on the state imposed by $p_{\text{BB84}}$ and $p_{\text{six-states}}$\label{sec:StateConstraints}}

\subsection{General parametrisation}

We consider measurements with settings $x,y$ and binary outcomes $a,b\in\{-1,+1\}$. From the knowledge that the measured systems are qubits, every measurement is \textit{a priori} a POVM with elements $\Pi_{x}^{a}\,=\,\gamma_{x}^{a}\Id+a\frac{\eta_{Ax}}{2}\sigma_{A_x}$ and $\Pi_{y}^{b}\,=\,\gamma_{y}^{b}\Id+b\frac{\eta_{By}}{2}\sigma_{B_y}$ with $\gamma^+_{x/y}+\gamma^-_{x/y}=1$ and $\sigma_{A_x/B_y}\equiv \hat{n}_{A_x/B_y}\cdot\vec{\sigma}$. Thus
\begin{equation}
P(a,b|x,y)=\gamma_x^a\gamma_y^b\,+\,a\,\frac{\eta_{Ax}}{2}\gamma_y^b \Tr{\rho \sigma_{A_x}\otimes\Id} \,+ \,b\,\gamma_x^a\frac{\eta_{By}}{2} \Tr{\rho \Id\otimes\sigma_{B_y}}\,+\,ab\,\frac{\eta_{Ax}\eta_{By}}{4} t_{A_xB_y}
\end{equation} with $\rho$ the state and $t_{A_xB_y}\equiv \Tr{\rho \sigma_{A_x}\otimes\sigma_{B_y}}$.

When the marginals are unbiased, i.e. $P(a|x)=P(b|y)=\frac{1}{2}$, we can work without loss of generality with states of the form
\begin{equation}
\rho=\frac{1}{4}(\Id\otimes\Id + m_A(y)\cdot \sigma_y \otimes \Id + m_B(y)\cdot \Id \otimes \sigma_y+ \sum_{i,j\in\{x,y,z\}}T_{ij}\cdot\sigma_i \otimes \sigma_j)
\end{equation}
and for the case of measurements with 3 settings, $m_A(y)=m_B(y)=0$. Note that the choice of $\sigma_y$ is completely arbitrary and the reason why $m_{A/B}(y)$ are left free (instead of $m_{A/B}(x)$ or $m_{A/B}(z)$)  is due to the parametrisation of the measurements in the following section.

 Now, given a state $\rho(\vec{m}_A,\vec{m}_B,T)$ with Bloch vectors $\vec{m}_{A/B}$ correlation matrix $T$, it's immediate that $\rho(-\vec{m}_A,-\vec{m}_B,T)$ is also a valid state, and so is their equal weighted mixture $\rho(0,0,T)$ which has the form:

\begin{equation}
\rho=\frac{1}{4}(\Id\otimes\Id + \sum_{i,j\in\{x,y,z\}}T_{ij}\cdot\sigma_i \otimes \sigma_j). \label{state}
\end{equation}

(of course, if the marginals were biased, such a state would not reproduce the observed statistics to start with). Indeed, the condition of unbiased marginals implies that all the observations are determined by the correlation matrix $T_{ij}=\Tr{\sigma_i\otimes\sigma_j\,\rho}$. Besides, if $\rho(\vec{m}_A,\vec{m}_B,T)$ is separable, then so are $\rho(-\vec{m}_A,-\vec{m}_B,T)$ and \textit{a fortiori} $\rho(0,0,T)$. Then, if all $\rho(0,0,T)$ that reproduce the observed statistics must be entangled, there cannot be any $\rho(\vec{m}_A,\vec{m}_B,T)$ that reproduces the statistics and is separable. Conversely, if one separable $\rho(0,0,T)$ does reproduce the observed statistics, then we know that the statistics can be reproduced by a separable state and there is no need to find others for our purpose.

\subsection{Case 1: six-state statistics}

The six-state statistics $p_{\text{six-states}}$ are given by
\begin{equation}
P(a,b|x,y)=\frac{1+(-1)^{\delta_{x,2}}ab\delta_{x,y}W}{4}\label{wernercorr}
\end{equation}
with $x,y\in\{0,1,2\}$. This immediately constrains all the POVMs to have $\gamma_{x/y}^+=\gamma_{x/y}^-=\frac{1}{2}$. It would also constrain the state to be such that $\Tr{\rho \sigma_{A_x}\otimes\Id}\,,=\,,\Tr{\rho \Id\otimes\sigma_{B_y}}\,,=\,,0$ for all the measurements, but we won't use this fact since we can directly use the form \eqref{state} of $\rho$ due to the fact that the marginals are unbiased.

Now we have to describe the constraints on the correlations $t_{A_xB_y}$. Since the state is left arbitrary, without loss of generality we can fix the measurement operators, $\hat{A}_x:=\sum_a a\Pi_x^{a}$ and $\hat{B}_y:=\sum_a b\Pi_y^{b}$, to be:
\begin{align}
\hat{A}_0=\eta_{A0}\sigma_z \;\;&\hat{B}_0 = \eta_{B0}\sigma_z\\
\hat{A}_1=\eta_{A1}(c_A\sigma_x+s_A\sigma_z) \;\;& \hat{B}_1=\eta_{B1}(c_B\sigma_x+s_B\sigma_z)\\
\hat{A}_2 = \eta_{A2}(D_A\sigma_y + E_A(d_A\sigma_x+e_A\sigma_z))\;\;& \hat{B}_2 = \eta_{B2}(D_B\sigma_y + E_B(d_B\sigma_x+e_B\sigma_z))
\end{align}
with $c_{A/B}^2+s_{A/B}^2=D_{A/B}^2+E_{A/B}^2=d_{A/B}^2+e_{A/B}^2=1$. Inserting the above expressions in \eqref{wernercorr}, we can obtain $T_{ij}= \Tr{\rho \sigma_{i}\otimes\sigma_{j}}$ for all $i,j\in\{x,y,z\}$:
\begin{align}
\eta_{A0}\eta_{B0}T_{zz}=W &\longrightarrow T_{zz}=\frac{W}{\eta_{A0}\eta_{B0}} \label{zz} \\
\eta_{A0}\eta_{B1}(c_BT_{zx}+s_BT_{zz})=0 &\longrightarrow T_{zx} = -\frac{s_B}{c_B}T_{zz} \label{zx} \\
\eta_{A1}\eta_{B0}(c_AT_{xz}+s_AT_{zz})=0 &\longrightarrow T_{xz} = -\frac{s_A}{c_A}T_{zz} \label{xz} \\
\begin{array}{r}\eta_{A1}\eta_{B1}(c_Ac_BT_{xx}+c_As_BT_{xz}\\+s_Ac_BT_{zx}+s_As_BT_{zz})=W 
\end{array}&\longrightarrow T_{xx}=\frac{1}{c_Ac_B}\left(\frac{W}{\eta_{A1}\eta_{B1}}+s_As_BT_{zz}\right) \label{xx} \\
\eta_{A0}\eta_{B2}(D_BT_{zy}+E_Bd_BT_{zx}+E_Be_BT_{zz})=0 &\longrightarrow T_{zy}=\frac{E_B}{D_B}(d_BT_{zx}+e_BT_{zz})\\
\eta_{A2}\eta_{B0}(D_AT_{yz}+E_Ad_AT_{xz}+E_Ae_AT_{zz})=0 &\longrightarrow T_{yz}=\frac{E_A}{D_A}(d_AT_{xz}+e_AT_{zz})\\
\begin{array}{r}\eta_{A1}\eta_{B2}\Big[ c_A(D_BT_{xy}+E_Bd_BT_{xx}+E_Be_BT_{xz})\\
+s_A(D_BT_{zy}+E_Bd_BT_{zx}+E_Be_BT_{zz})\Big]=0 
\end{array} &\longrightarrow 
\begin{array}{r}
T_{xy}=-\frac{s_A}{c_A}T_{zy}-\frac{s_A}{c_A}\frac{E_B}{D_B}(d_BT_{zx}\\+e_BT_{zz})-\frac{E_B}{D_B}(d_BT_{xx}+e_BT_{xz})
\end{array}\\
\begin{array}{r}\eta_{A2}\eta_{B1}\Big[c_B(D_AT_{yx}+E_Ad_AT_{xx}+E_Ae_AT_{zx})\\
+s_B(D_AT_{yz}+E_Ad_AT_{xz}+E_Ae_AT_{zz})\Big]=0 
\end{array} &\longrightarrow
\begin{array}{r}
T_{yx}=-\frac{s_B}{c_B}T_{yz}-\frac{s_B}{c_B}\frac{E_A}{D_A}(d_AT_{xz}\\+e_AT_{zz})-\frac{E_A}{D_A}(d_AT_{xx}+e_AT_{zx})
\end{array}\\
\begin{array}{r}\eta_{A2}\eta_{B2}\Big[ D_AD_BT_{yy} + D_AE_B(d_BT_{yx}+e_BT_{yz})\\+ E_AD_B(d_AT_{xy}+e_AT_{zy})\\+E_AE_B(d_Ad_BT_{xx}+d_Ae_BT_{xz}\\+e_Ad_BT_{zx}+e_Ae_BT_{zz})\Big]=-W \end{array} &\longrightarrow \begin{array}{r}T_{yy}=-\frac{1}{D_AD_B}\Big[(\frac{W}{\eta_{A2}\eta_{B2}} + \\D_AE_B(d_BT_{yx}+e_BT_{yz})\\+ E_AD_B(d_AT_{xy}+e_AT_{zy})\\+E_AE_B(d_Ad_BT_{xx}+d_Ae_BT_{xz}\\+e_Ad_BT_{zx}+e_Ae_BT_{zz})\Big]\,. \label{yy}
\end{array}
\end{align}

\subsection{Case 2: BB84 statistics}

For the case where the measurements have two settings $x,y\in\{0,1\}$ and two outcomes $a,b\in\{-1,+1\}$ and the observed statistics correspond to $p_{\text{BB84}}$, given explicitly by:
\begin{equation}
P(a,b|x,y)=\frac{1+ab\delta_{x,y}W}{4}
\end{equation}
The marginals are unbiased, so we can again work with a state of the form \eqref{state}. Also, the values of $T_{zz}$, $T_{zx}$, $T_{xz}$ and $T_{xx}$ are constrained according by the same equations as above, \eqref{zz}, \eqref{zx}, \eqref{xz} and \eqref{xx} respectively. However, the other $T_{ij}$ are now constrained only by the requirement that $\rho\geq 0$.

\subsection{Ideal statistics ($W=1$)}

If $W=1$, \eqref{zz} becomes $T_{zz}=\frac{1}{\eta_{A0}\eta_{B0}}$. But $|T_{zz}|\leq 1$: this implies immediately $\eta_{A0}=\eta_{B0}=1$ and $T_{zz}=1$. In turn, this means that $T_{zx}=T_{xz}=T_{zy}=T_{yz}=0$. Indeed, if $T_{zx}=\lambda$, then $\Tr{\rho \sigma_z\otimes (\cos\theta\sigma_z+\sin\theta\sigma_x)}=\cos\theta+\lambda\sin\theta$, which is always larger than 1 for $\theta$ small enough if $\lambda\neq 0$.

Then, the condition \eqref{xx} reads $T_{xx}=\frac{1}{\eta_{A1}\eta_{B1}}$, so as above we find $\eta_{A1}=\eta_{B1}=1$ and $T_{xx}=1$, and consequently $T_{xy}=T_{yx}=0$. Finally, the equations $Tr(\rho \sigma_z\otimes\sigma_z)=Tr(\rho \sigma_x\otimes\sigma_x)=1$ identify uniquely $\rho=\ket{\Phi^+}\bra{\Phi^+}$, i.e. $T_{yy}=-1$.

In conclusion: under the knowledge that the systems are qubits, the ideal case $W=1$ of both the BB84 and the six-state statistics provide a self-testing of the state $\ket{\Phi^+}$ and the projective measurements $A_0=B_0=\sigma_z$, $A_1=B_1=\sigma_x$, and for the six-state case $A_2=B_2=\sigma_y$.



\section{Supplementary Information: BB84 protocol with imperfect detectors efficiency}\label{sec:cryptoAppendix}
We consider the BB84 protocol in the case where the detectors efficiencies, $\epsilon_A$ and $\epsilon_B$, are not perfect. Also, we study a possible strategy that the outcomes, $a$ and $b$, are printed as ``+1'' when the detector does not click. Moreover, for simplicity, we fix $\epsilon_A=\epsilon_B$. Hence, we arrive at the following observed correlations:
\begin{align}
P(a=0,b=0|x,y)&=\epsilon^2\left(\frac{1+\delta_{x,y}W}{4}\right)+\epsilon(1-\epsilon)+(1-\epsilon)^2 \label{noisyeq1} \\
P(a=0,b=1|x,y)&=\epsilon^2\left(\frac{1-\delta_{x,y}W}{4}\right)+\frac{\epsilon(1-\epsilon)}{2} \\
P(a=1,b=0|x,y)&=\epsilon^2\left(\frac{1-\delta_{x,y}W}{4}\right)+\frac{\epsilon(1-\epsilon)}{2} \\
P(a=1,b=1|x,y)&=\epsilon^2\left(\frac{1+\delta_{x,y}W}{4}\right) \label{noisyeq4}
\end{align}
Notice that the Quantum Bit Error Rate, $Q$, is given by:
\begin{equation}
Q=\epsilon^2\left(\frac{1-\delta_{xy}W}{2}\right)+\epsilon(1-\epsilon)
\end{equation}
Also, in this case, the marginals are no longer unbiased and is given by $P(a/b=+1|x/y)-P(a/b=-1|x/y)=1-\epsilon$. Hence, the parametrisation of the measurements and state in the previous sections are no longer valid. Here, we have to consider all possible bipartite qubits states given by:
\begin{equation}
\rho=\frac{1}{4}(\Id\otimes\Id + \vec{m}_A\cdot \vec{\sigma} \otimes \Id +  \Id \otimes \vec{m}_B \cdot \vec{\sigma}+ \sum_{i,j\in\{x,y,z\}}T_{ij}\cdot\sigma_i \otimes \sigma_j)
\end{equation}
Since the state is left free, without loss of generality, we can consider measurements with POVM elements of the form:
\begin{align}
\Pi_{x=0}^{a=+1}&=\alpha_1\Id + \beta_1\sigma_z\\
\Pi_{x=1}^{a=+1}&=\alpha_2\Id + \beta_2\sigma_z + \beta_3\sigma_x\\
\Pi_{y=0}^{b=+1}&=\gamma_1\Id + \delta_1\sigma_z\\
\Pi_{y=1}^{b=+1}&=\gamma_2\Id + \delta_2\sigma_z + \delta_3\sigma_x\\
\Pi_{x/y}^{a/b=-1}&=\Id-\Pi_{x/y}^{a/b=+1}
\end{align}
The only additional requirement to ensure the validity of the measurements is that each POVM elements defined above must be constrained to be positive semi-definite. 

With the state and the measurements well defined, we can now write down the constraints on the state given the observed correlations given in equations \eqref{noisyeq1} to \eqref{noisyeq4}. Using the same method employed in previous section, we arrive at the following constraints:
\begin{align}
m_A(z)&=\frac{2-2\alpha_1-\epsilon}{2\beta_1}\\
m_B(z)&=\frac{2-2\gamma_1-\epsilon}{2\delta_1}\\
m_A(x)&=\frac{2-2\alpha_2-\epsilon-2\beta_2m_A(z)}{2\beta_3}\\
m_B(x)&=\frac{2-2\gamma_2-\epsilon-2\delta_2m_A(z)}{2\delta_3}\\
T_{zz}&=\frac{1}{4\beta_1\delta_1}(\epsilon^2W+(1-\epsilon)^2\\
&-(2\alpha_1-1)(2\gamma_1-1)-2(2\alpha_1-1)\delta_1m_B(z)\nonumber\\
&-2(2\gamma_1-1)\beta_1m_A(z))\nonumber\\
T_{zx}&=\frac{1}{4\beta_1\delta_3}((1-\epsilon)^2-(2\alpha_1-1)(2\gamma_2-1)\\
&-2(2\alpha_1-1)\delta_2m_B(z)-2(2\alpha_1-1)\delta_3m_B(x)\nonumber\\
&-2\beta_1(2\gamma_2-1)m_B(z)-4\beta_1\delta_2T_{zz})\nonumber\\
T_{xz}&=\frac{1}{4\delta_1\beta_3}((1-\epsilon)^2-(2\gamma_1-1)(2\alpha_2-1)\\
&-2(2\gamma_1-1)\beta_2m_A(z)-2(2\gamma_1-1)\beta_3m_A(x)\nonumber\\
&-2\delta_1(2\alpha_2-1)m_A(z)-4\delta_1\beta_2T_{zz})\nonumber\\
T_{xx}&=\frac{1}{4\beta_3\delta_3}(\epsilon^2W+(1-\epsilon)^2\\
&-(2\alpha_2-1)(2\gamma_2-1)-2(2\alpha_2-1)\delta_2m_B(z)\nonumber\\
&-2(2\alpha_2-1)\delta_3m_B(x)-2(2\gamma_2-1)\beta_2m_A(z)\nonumber\\
&-2(2\gamma_2-1)\beta_3m_A(x)-4\beta_2\delta_2T_{zz}\nonumber\\
&-4\beta_2\delta_3T_{zx}-4\beta_3\delta_2T_{xz})\nonumber
\end{align}
Hence, we can perform the minimisation of the function $1-h(Q)-\chi(A:E)$ over the free parameters; $\alpha_{1/2}$, $\beta_{1/2/3}$, $\gamma_{1/2}$, $\delta_{1/2/3}$, $m_{A/B}(y)$, $T_{xy}$, $T_{yy}$, $T_{zy}$, $T_{yx}$ and $T_{yz}$ such that the POVM elements and the state are positive semi-definite with different values of $W$ and $\epsilon$. The optimal result of the minimisation result will give the secret key rate for any particular values of observed $Q$ for a particular detectors efficiencies of the experimental setup, $\epsilon$. 

\begin{figure}
  \subfloat{
    \includegraphics[width=0.46\textwidth]{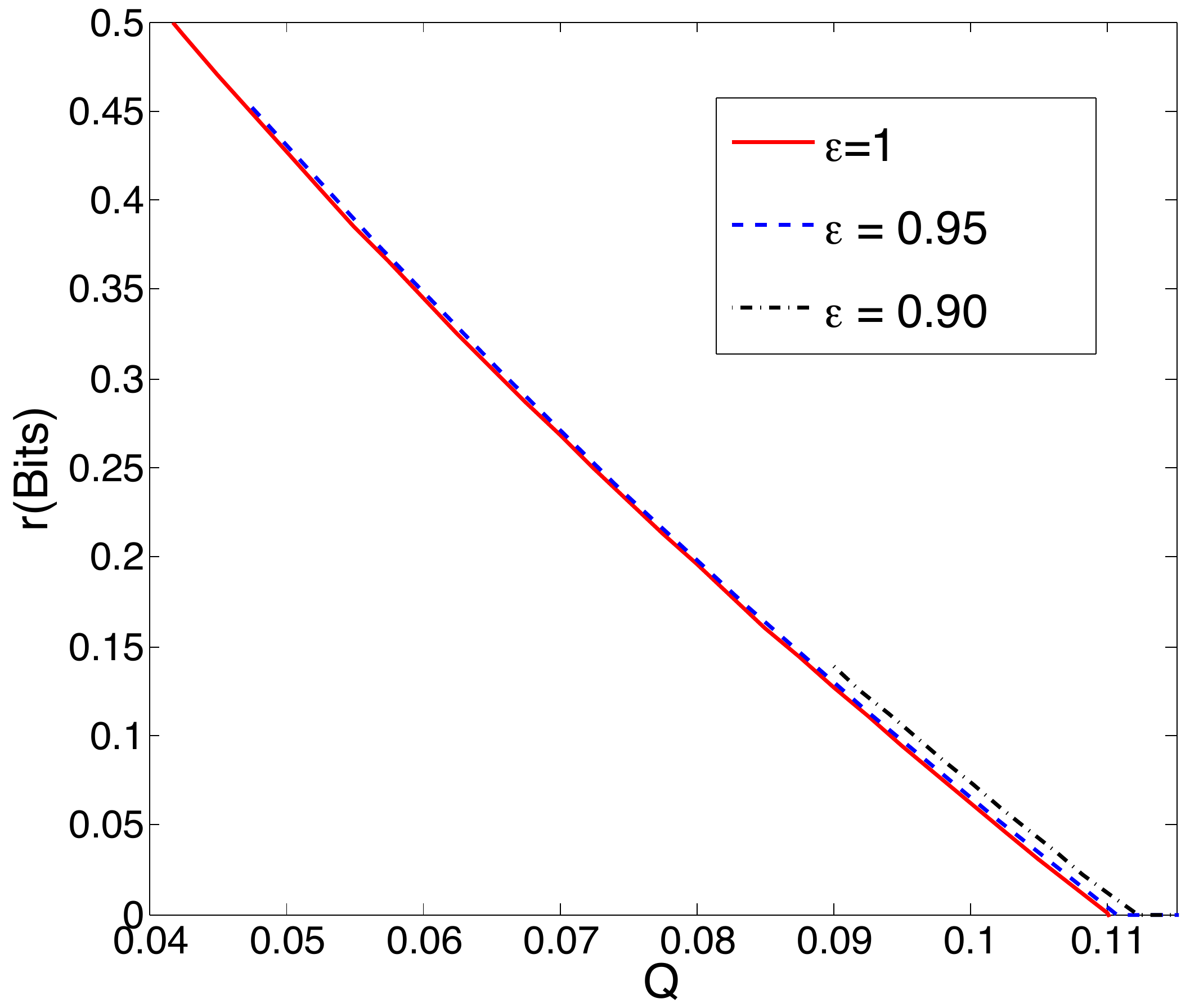}}\quad
 \subfloat{\includegraphics[width=0.46\textwidth]{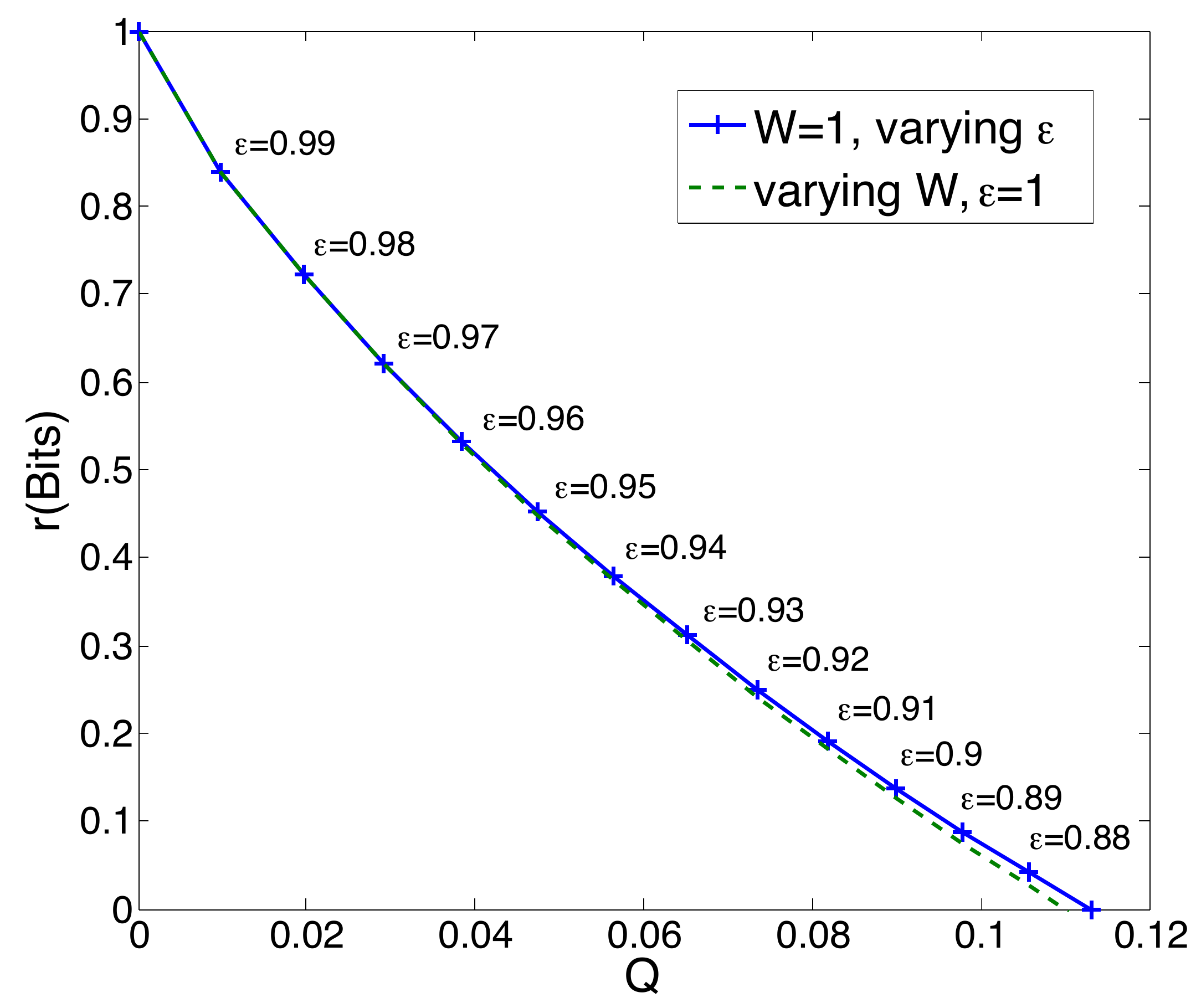}}
 \caption{Optimisation results for secret key rates for BB84 protocol with imperfect detectors efficiencies. \textbf{(Left)} Plot of secret key rate against $Q$ for different detectors efficiencies; $\epsilon=1$ (red solid line), $\epsilon=0.95$ (blue dashed line) and $\epsilon=0.90$ (black dashed dotted line) \textbf{(Right)} Plot of secret key rate against $Q$ with $W=1$ and varying $\epsilon$ (blue solid line with ``+'' data points), this curve represents the highest achievable secret key rate for a given $\epsilon$. The secret key rate of $\epsilon=1$ (green dashed line) is also plotted on the same graph for comparison. }
 \label{noisybb84}
\end{figure}

Under such analysis, we assume that an Eavesdropper Eve does not have the information of when the detectors fail to click and she is not allowed cause loss in the quantum channel between Alice and Bob i.e. Eve's only resource is the purification of the joint state between Alice and Bob. Even with such optimism, the result in FIG. \ref{noisybb84} (right) shows that if the detectors have efficiencies, $\epsilon\leq 0.87$, no secret keys can be established between Alice and Bob via BB84 protocol assuming the signal states are qubits. 

Notice that in FIG. \ref{noisybb84} (left), the results suggest that a higher secret key rate can be obtained with lower detector efficiency for a given value of $Q$. This is due the assumption that Eve does not have the knowledge and control over the events of ``no detection''. This implies that a contribution of $Q$ is purely due to these events of ``no detection'' which do not give Eve any information on Alice's keys. Taking this into account, for a given (or a lack of) attack by Eve, a higher value of $\epsilon$ gives a lower secret key rate, as shown in FIG. \ref{noisybb84} (right).
\end{document}